**On the Relationship between Structure and Dynamics in a Supercooled Liquid**

Asaph Widmer-Cooper and Peter Harrowell
School of Chemistry, University of Sydney, Sydney 2006, New South Wales, Australia

Abstract

We present the dynamic propensity distribution as an explicit measure of the degree to which the dynamics in a liquid over the time scale of structural relaxation is determined by the initial configuration. We then examine, for a binary mixture of soft disks in 2D, the correlation between the spatial distribution of propensity and that of two local measures of configuration structure: the local composition and local free volume. While the small particles dominate the high propensity population, we find no strong correlation between either the local composition or the local free volume and the propensity. It is argued that this is a generic failure of purely local structural measures to capture the inherently non-local character of collective behaviour.

## 1. Introduction

The changing character of relaxation and transport dynamics in a liquid as it is supercooled is the result of the increasing configurational restrictions exerted by the particle interactions. Structure, in other words, plays an increasingly complex role in determining how and where particle motion will occur. In this paper we approach the problem through simulations of a simple glass-forming liquid, a 2D binary mixture of soft disks. The great advantage of this system is that much of the increasing complexity of the collective motion, seen on supercooling, can be directly visualised.

The paper is organised as follows. In the following two Sections we present a brief review of the treatment of dynamics in liquids and in solids, followed by a specific consideration of the description of collective dynamics in simple dense liquids and the central role played by dynamic heterogeneity. In Section 4 we look at an explicit quantification of the degree to which particle dynamics is determined by a given configuration with the introduction of a quantity we call the dynamic propensity of a configuration. We then test two reduced descriptions of the configuration, one based on the local composition and the other on the local free volume, to see if either of these local measures of the structure show a strong enough correlation with the propensity to correspond to a causal link. Neither do. We discuss possible nonlocal extensions of the free volume that might better identify this kinetic structure.

## 2. Dynamics in Liquids and Solids: an Overview

Self diffusion in a simple liquid above its melting point is generally well described by mean field theory - whether one looks at the problem via kinetic theory or a generalised Langevin with memory [1] (the latter including the mode coupling treatments of the memory function [2]). These theories are mean field in the sense that the dynamics are determined by the average structure of the liquid. In contrast, diffusion in solids is dominated by rare fluctuations in the structure - point defects, dislocations and grain boundaries. As happens whenever kinetics are subject to rare events (nucleation phenomena or fracture, for example), the associated theory treats the relevant fluctuations explicitly rather than trust to the dubious accuracy of the wings of distributions. This is certainly the case in the extensive theoretical literature concerning the defect and grain-boundary mediated transport in solids [3]. Cahn [4] has argued that, in general, it is this strong link between microscopic structure and physical properties that essentially defines an established field of material science.

The continuous transition from fluid to solid associated with the glass transition traverses between these two extremes. The associated conceptual transition from mean field to fluctuation-dominated is, perhaps, not immediately evident from the literature. The major theoretical treatment of the glass transition - the mode coupling theory (MCT) [2] - incorporates the average liquid structure through vertex functions. This qualifies the MCT as a mean field theory. We note, however, that the term 'mean field' does require some qualification. A hierarchy of generalised Langevin theories can be can be imagined in which the neglect of fluctuations (the 'mean field' approximation) occurs at increasingly higher orders of correlations. Szamel [5] has recently presented a mode coupling theory of relaxation in a simple lattice model of a glass in which the factorization is applied to one order higher in correlation to that of the standard approximation. This theory captures scaling laws previously thought to be obtainable only from an explicit treatment of the rare fluctuations responsible for dynamics [6].

In spite of its mean field character, the evidence that the mode coupling theory can provide a quantitative treatment of diffusion and structural relaxation leading up to the glass transition is impressive [2]. The more recent success of mode coupling theory in providing a unified treatment of colloidal glasses and associating gels is quite remarkable [7]. The problem is that the transition itself - the ideal glass transition - is an artifact of the mathematical structure of the self-consistency introduced by the factorisation approximation. From one point of view, this is an attractive feature of the model - the fact that arrest enters naturally without having to burden the treatment with all the physical correlations that actually stabilise the solid. On the question of the actual physical origins of rigidity of the glass, however, the mode coupling theory is silent. It is necessary to look elsewhere to understand the relationship between structure and dynamics in the liquid as this rigid state is approached

## 3. Cooperativity and the Spatial Distribution of Dynamics

If one was to attempt as detailed a characterisation of dynamics in a solid as one could, *without* making use of explicit structural descriptions, then the outcome would probably consist of a description of the long lived spatial heterogeneity in the particle motion.

While a handful of particles would be found to undergo substantial motion, the majority would simply vibrate around a mean position. The appreciation of exactly this type of characterisation as applied to supercooled liquids, along with the consequences of these heterogeneous dynamics, represents one of the major developments in the last 15 years in the study of glass-forming liquids [8].

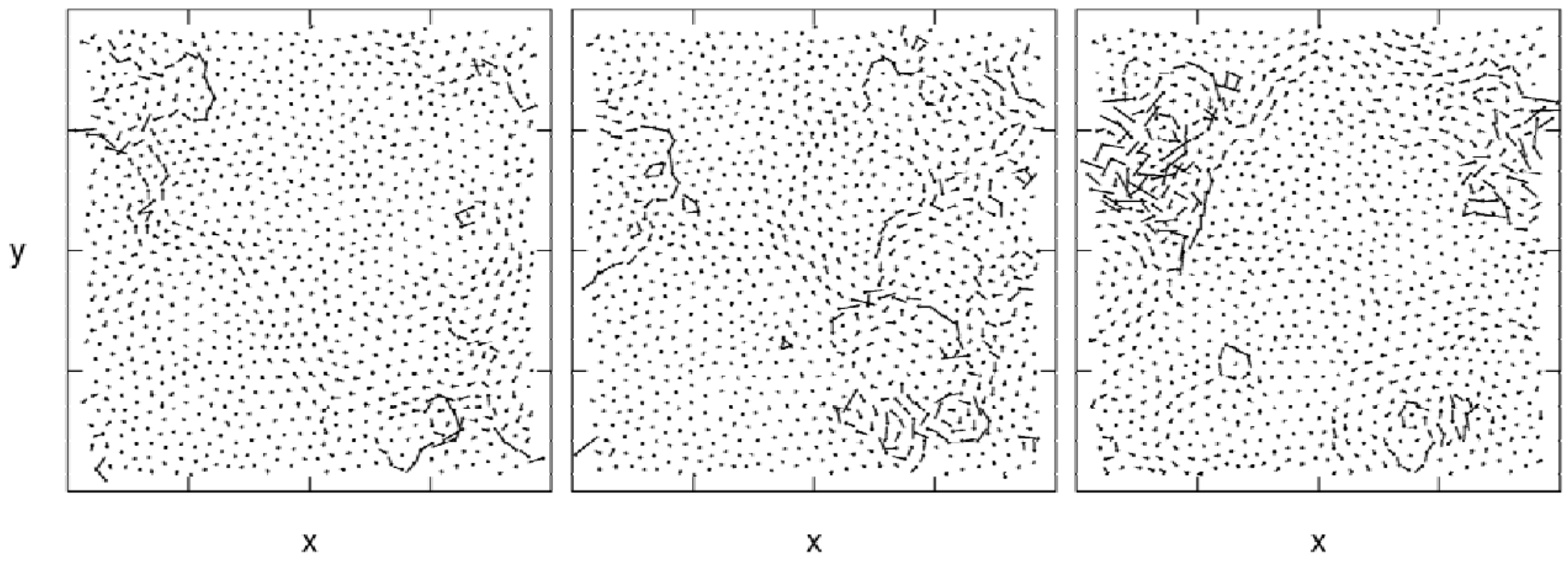

Figure 1. Particle displacements over $1.5\tau_e$ (see text) for the 2D equimolar binary mixture at a reduced temperature of 0.4 . The three trajectories correspond to runs from the same initial configuration but with randomly assigned initial momenta.

There at least two ways of analysing the spatial pattern of particle dynamics such as that shown in Fig.1. One approach is to consider the correlation between the initial particle configuration and the subsequent dynamics. This approach, the subject of this paper, will be considered in the Sections to follow. The second is to consider the correlation between the mobile particles themselves. It is this latter approach that has received considerable attention and is the subject of the remainder of this Section.

Particle motion in dense liquids is, to a large extent, entrained so that particles follow along in each others path. In 1998, Donati et al [9] showed that displacements in a supercooled liquid exhibited a strong tendency to locally align. Similar 'strings' of displacements are seen in Fig.1. The dynamics associated with such correlations are generally quite complex. To appreciate this, consider, first, the *simple* scenario of a diffusing vacancy. The motion of such a defect will leave a linear trail of particle displacements to mark its passage. The linear character of this pattern of displacements reflects a rough local conservation of free volume, i.e. the volume left free by the motion of a particle is more likely to be filled by a single particle rather than the collective rearrangement of a number of particles. There is also a correspondence between spatial structure and temporal sequence so that one end of the resulting string of displacements represents the first step while the other end represents the last step. Ritort and Sollich [6] have recently reviewed the predictions of number of diffusing defect models.

The diffusing defect picture, presented above, ignores the possibility that the propensity for motion lies distributed in a configuration and that relaxation is not a simple consequence of the transport of a rare fluctuation (even one more complex than a simple vacancy) but rather a sequence of unlocking events which adds up, over the observation interval, to a linear path. Vogel et al [10] have presented simulation evidence of just this latter process. Delaye and Limoge [11], in an interesting study, considered the different fates of vacancies created in a model glass. The resulting behaviour was divided into 3 groups: those defects that remained stable and stationary, those that relaxed by propagation (the diffusing defects) and the third group that relaxed by being, essentially, 'absorbed' back into the surrounding disordered medium through a local collective rearrangement. The presence of this last process distinguishes the amorphous material from the crystalline. In terms of a simple model that can capture this more complex collective behavior, the class of facilitated spin flip models introduced by Fredrickson and Andersen [12] and extended by Jäckle [13] are particularly useful. The term 'facilitated' refers to the idea that the local state of a system can influence the kinetics of adjacent regions. To date, models of facilitated dynamics have all relied on the introduction of explicit kinetic constraints. Recently, Garrahan and Chandler [14] have proposed that this idea of 'facilitation' represents a general aspect of dense disordered phases. The interesting unanswered question here is whether the implied general mapping from systems of interacting particles to systems governed by kinetic constraints exists. Central to this question is the need to understand the degree to which a particle configuration determines the propensity of particles to subsequently move. This is the subject of this paper.

The analysis of the correlation between particle displacements sketched here provides i) a compact summary of the information represented by dynamic heterogeneities, ii) explanation of some observed features of relaxation functions and transport behaviour in terms of microscopic dynamics and iii) the prospect of identifying kinetic rules that govern relaxation in systems. This approach, however, does not explain what it is about a configuration that permits motion in one region but not in another nor how this distribution varies with temperature, composition, particle interactions, etc. One could imagine, for example, studying transport in crystalline solids via this description, amassing a considerable amount of phenomenological information about the dynamic heterogeneities and yet never arrive at a clear structural (and, hence, predictive) picture of vacancies and interstitials. For this reason, we would like to address directly the question of the relationship between structure and dynamics.

## 4. The Spatial Distribution of Dynamic Propensity: Refining the Structure→Dynamics Link

To the degree that the study of transport in supercooled liquids and glasses aspires to join the ranks of established material science, as identified in ref. 4, its goal is to establish useful atomistic descriptions of the collective processes that govern dynamics. Structure, in this context, refers to a reduced measure of a particle configuration that still retains the information on which the subsequent dynamics depends.

Before we approach the question of how structure determines dynamics, we must first establish exactly what aspects of the dynamics of a disordered system can be explicitly correlated with a given configuration. The degree to which the liquid dynamics reflects a persistent influence of a configuration is related to the idea, introduced at the start of this paper, of the crossover between liquid- and solid-like behaviour on cooling. One measure of this transition from the liquid to solid-like descriptions is the crossover temperature proposed by Goldstein in 1969 [15]. The crossover temperature marks the transition from the high temperature liquid where momentum transfer (as binary collisions and, collectively, in hydrodynamic modes) plays a dominant role to the low temperature liquid in which dynamics is governed by activated transitions from one stable configuration to another. The dynamics is said [16] to become 'landscape-dominated', the landscape referring to the potential energy surface over the configuration space. The configuration space, in other words, has begun to break up into kinetically isolated domains.

We propose to answer the question - what aspect of particle dynamics in a liquid is determined by a given configuration? - with the following construction [17]. Consider an ensemble of $N$-particle trajectories, all starting from the same configuration but with momenta assigned randomly from the appropriate Maxwell-Boltzmann distribution. The squared displacement of each particle over a fixed time interval can be averaged over this ensemble of trajectories. (We will call this an *iso-configurational* ensemble in reference to the common initial configuration.) The time interval needs to be chosen to permit the observation of dynamic heterogeneities. Here we have chosen the interval to be 1.5 x $\tau_e$ where $\tau_e$, the structural relaxation time, is defined as the time at which the intermediate scattering function, measured at the wave vector of the first peak in the structure factor, has decayed to 1/e . If the initial configuration had no influence on the particle displacements then we would expect to see no spatial variation in the ensemble averaged squared displacements of particles. If, on the other hand, the dynamics over the chosen time interval was strongly dependent upon the initial configuration then we would expect to see significant spatial variations in the mean squared displacements. This ensemble mean squared displacement does not correspond to the actual squared displacement of the particle in any particular run but, rather, reflects the particle's propensity for displacement. We shall call this ensemble mean squared displacement the *dynamic propensity* of the particle.[17]

We have calculated the propensity for a binary mixture of soft disks in 2D. The model consists of an equimolar mixture of particles interacting via a repulsive potential of the form $\varphi_{ij}(r) = 4\,\varepsilon(\sigma_{ij}/r)^{12}$ where $\sigma_{AA} = 1.4$, $\sigma_{BB} = 1.0$ and $\sigma_{AB} = 1.2$. This glass-forming mixture has been extensively characterised and details of the model and the molecular dynamics algorithm can be found in ref. [17]. In Figs. 2 and 3 we plot contour maps of the propensity of individual configurations at reduced temperatures of T = 1.0 and 0.4, respectively. (The contour maps are generated by interpolating the particle propensities onto a grid (using a modification of Shepards method as implemented in the NAG libraries). Care has been taken to ensure that the interpolation has not introduced any spurious features.) The structural relaxation time at T = 0.4 is roughly 1000 times that at the higher temperature. Note the difference in both the spatial distribution of propensities and the variation in the amplitude. These maps provide an explicit real space expression of the crossover in the dynamics. The striking structure of the low temperature map corresponds to the aspect of the dynamics that is governed by the selected configuration.

We shall now consider a number of reduced descriptions of the initial configuration and see how well these measures correlate with the observed propensities.

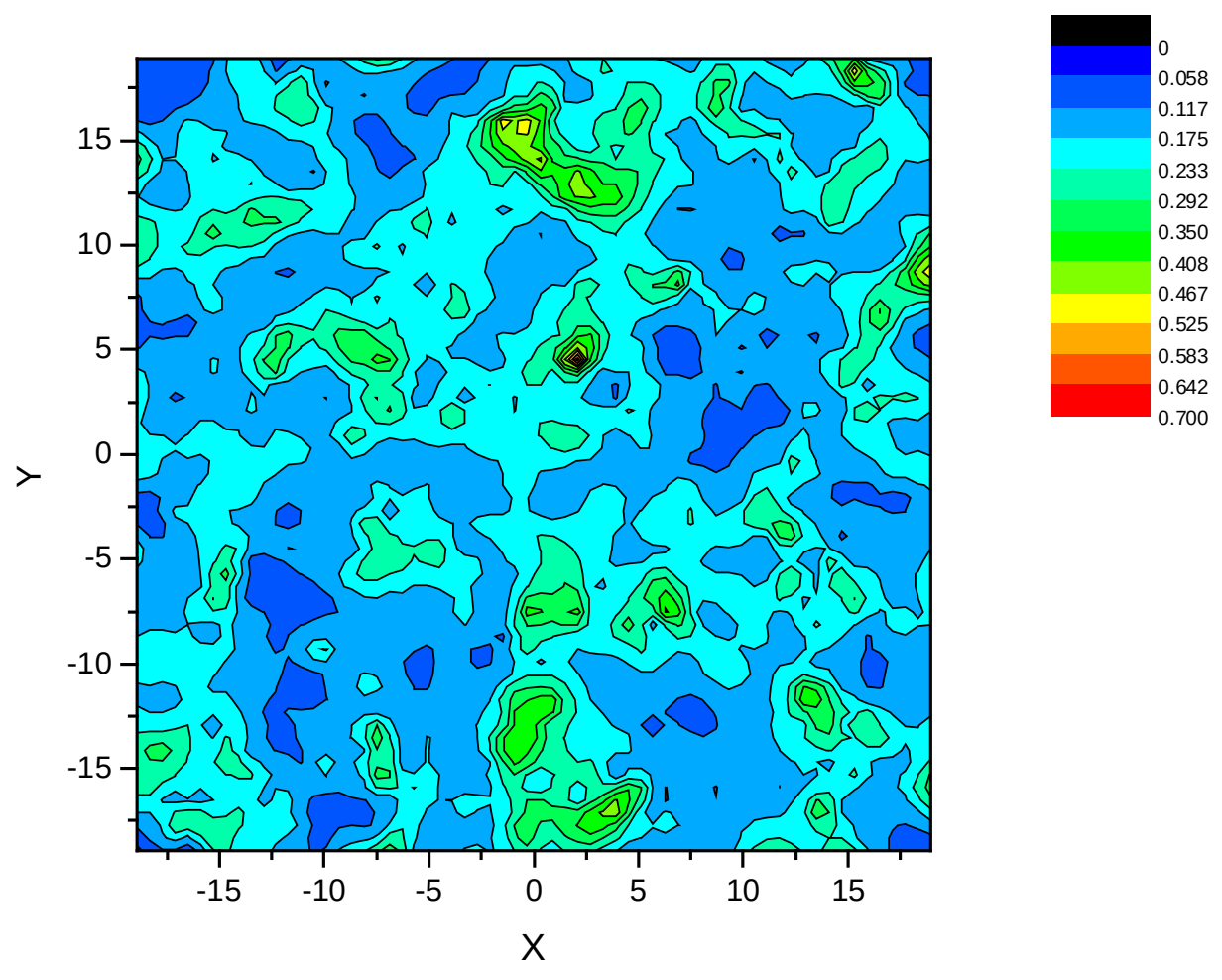

Figure 2. The propensity for a configuration at T = 1.0. All propensity maps represent averages over 100 runs.

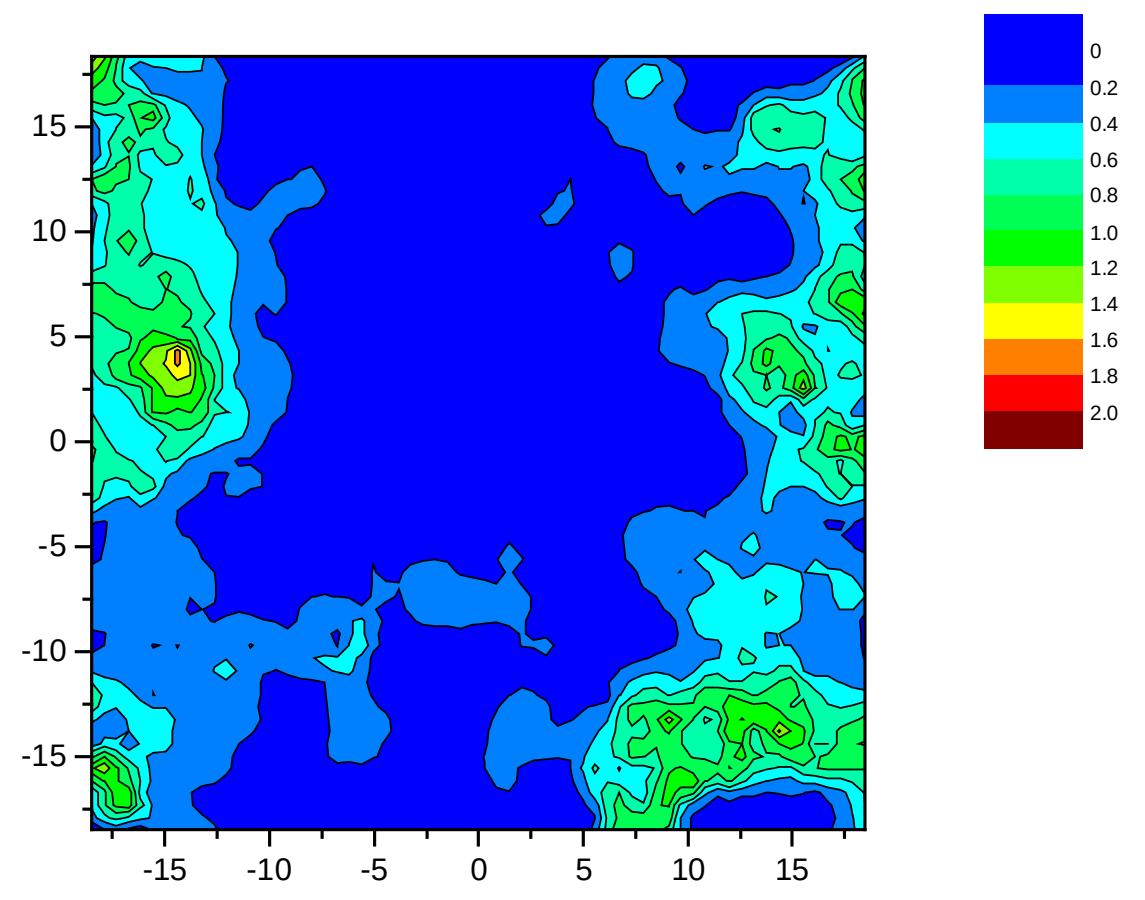

Figure 3. The propensity for a configuration at T = 0.4.

We shall first consider a measure of the local environment of each particle in terms of the number of neighbours it has of each type. This measure would be expected to reflect other possible local measures (local topology, potential energy per particle (but local PE is related to free volume), etc.) and so represents a general approach to structure that we might term 'chemically oriented'. We have identified a particular neighborhood with the following notation. A small particle with *m* small neighbours and *n* large neighbours is designated as S*mn* and the analogous large particle is indicated as L*mn*. In Fig. 4b we present the populations of the different environments in our 2D binary mixture at T = 0.4. Note that the small particle find themselves in either 5- or 6-fold environments while the large particles have either 6 or 7 neighbours, hence the 4 distinct clumps in the distribution. Also note the large number of different environments.

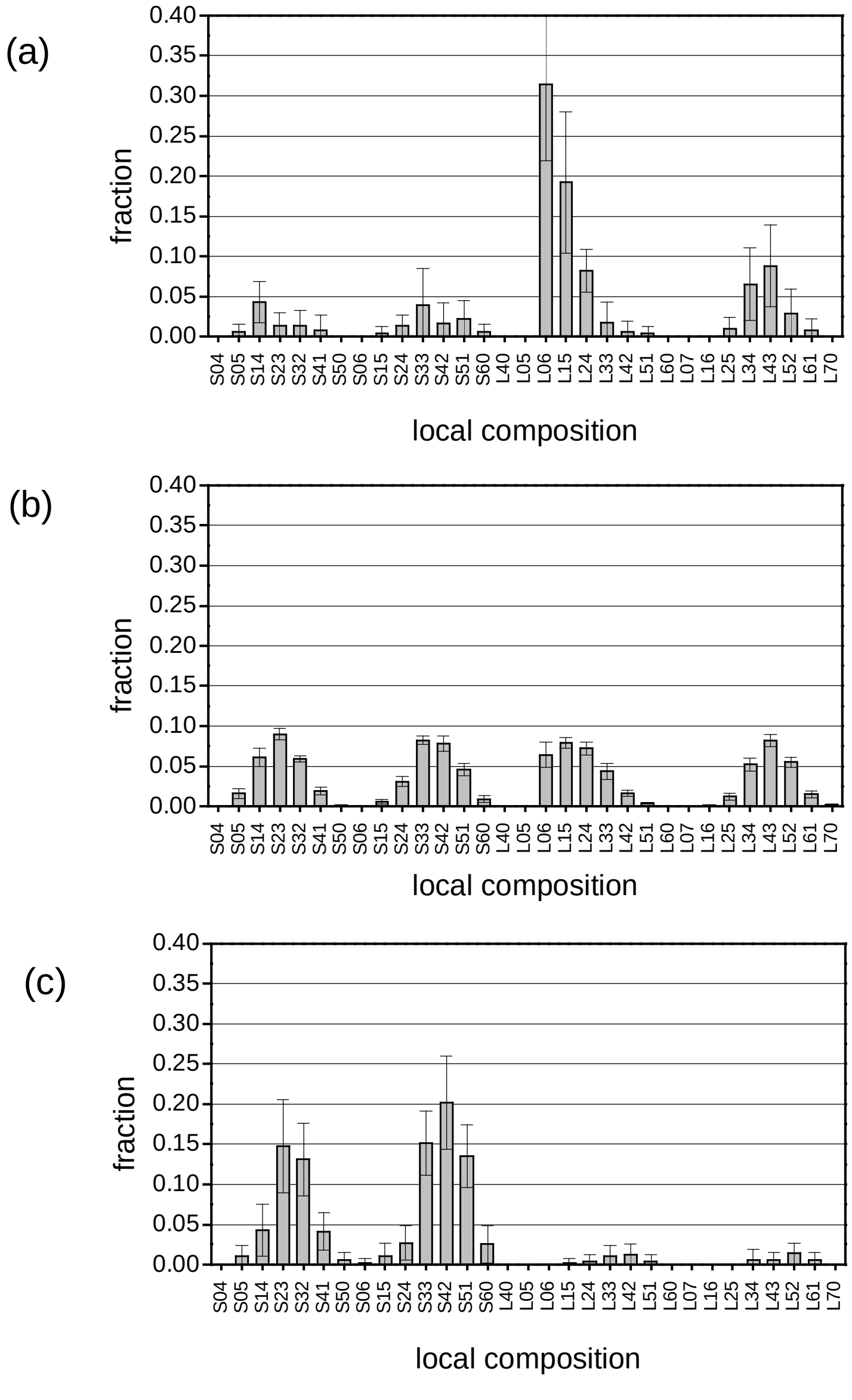

Figure 4. The distribution of local compositions in the binary mixture at T = 0.4 for a) the slowest 5% of particles, b) all the particles and c) the fastest 5% of particles. The composition code is explained in the text.

Unlike other model binary glass formers, our purely steric interactions do not lead to any significant chemical ordering in the liquid. In Fig. 4a we have presented the distribution of environments found among the slowest 5% of particles (i.e. the lowest 5% when ranked in terms of their propensity). This group is clearly dominated by large particles, in particular the L06 arrangement, corresponding to hexagonal packing of large particles. While this correlation is interesting, our real interest concerns places where something *will* happen. In Fig. 4c we look at the environments of the 5% fastest particles (i.e. the highest 5% when ranked in terms of their propensity). We see a clear dominance of the small particles. The distribution of environments among the small particles, however, appears to be similar to that of the total distribution. This implies that there is no special local environment that produces high propensity and that some property other than local composition must be responsible for determining which of the small particles exhibits high propensity.

Next we turn to a local geometric measure, the free volume. There is a long history of using the average total free volume as means of relating a thermodynamic property to the diffusion constant in a liquid. Success of these free volume theories is measured by how well the empirical relationship between average total free volume and the diffusion constant can be described by a simple functional form. It is not uncommon for this success to be interpreted as implying that a strong correlation exists between the local free volume and the microscopic dynamics. We shall test this proposition in the case of the 2D soft disk mixture.

We define the free volume of a particle as the area accessible to that particle with all its neighbors fixed. For the purpose of this calculation, the particles are assumed to be hard disks with a contact distance given by 0.9 $\sigma_{ij}$ where j is the identity of the central particle. (We find that varying these lengths from 0.85 $\sigma_{ij}$ to 1.0 $\sigma_{ij}$, while clearly altering the magnitude of the free volume, does not significantly change the ordering of the particles in a configuration in terms of the calculated free volume.) Our free volume calculations involved an algorithm similar to that described by Sastry et al [18]. To allow for direct comparison we present, in Fig. 5, a contour plot of the propensity on which the positions of the 10% of particles with the largest relative free volumes are indicated.

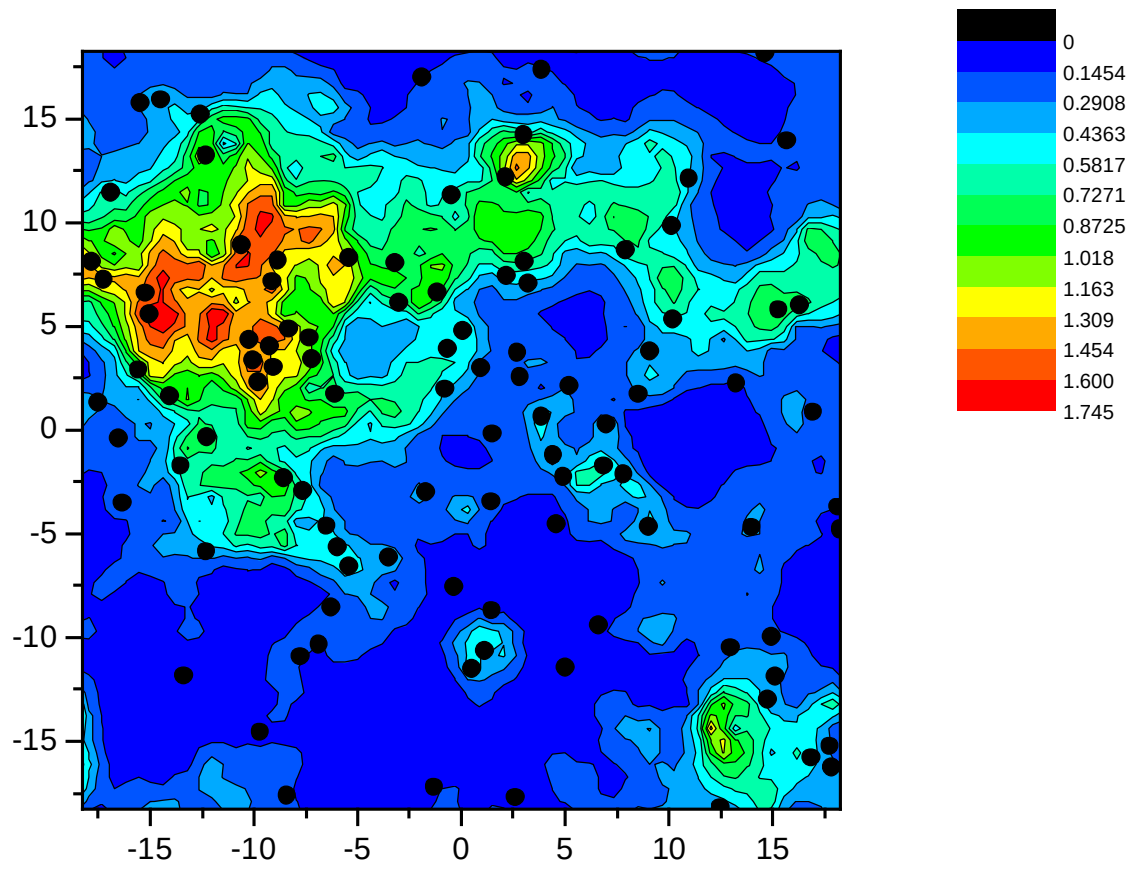

Figure 5. Spatial map of propensity for a configuration at T = 0.4. Black dots indicate the 10% of particles with the highest relative free volume.

(The relative free volume is the free volume of a particle *i* divided by the particle volume calculated using the effective hard disc radius associated with the interaction between two particles of the same species as particle *i*.)

In spite of the coincidence of high propensity and high relative free volume in some cases, there are clearly many 'false positives'. In Fig. 6 we plot the scatter plot of propensity against the relative free volume for the same configuration shown in Fig.5.

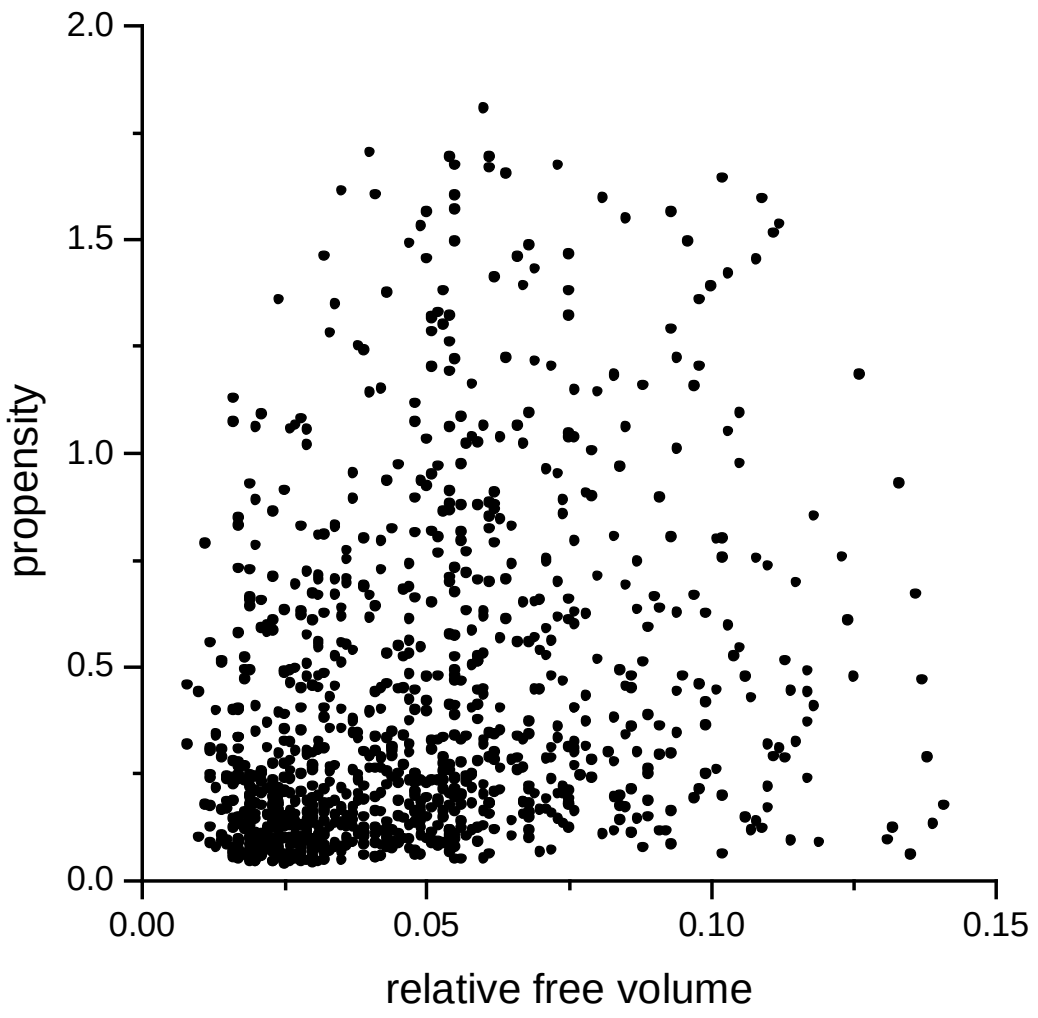

Figure 6. A scatter plot of propensity vs relative free volume for the same configuration as used in Fig.5. We find no significant correlation is present.

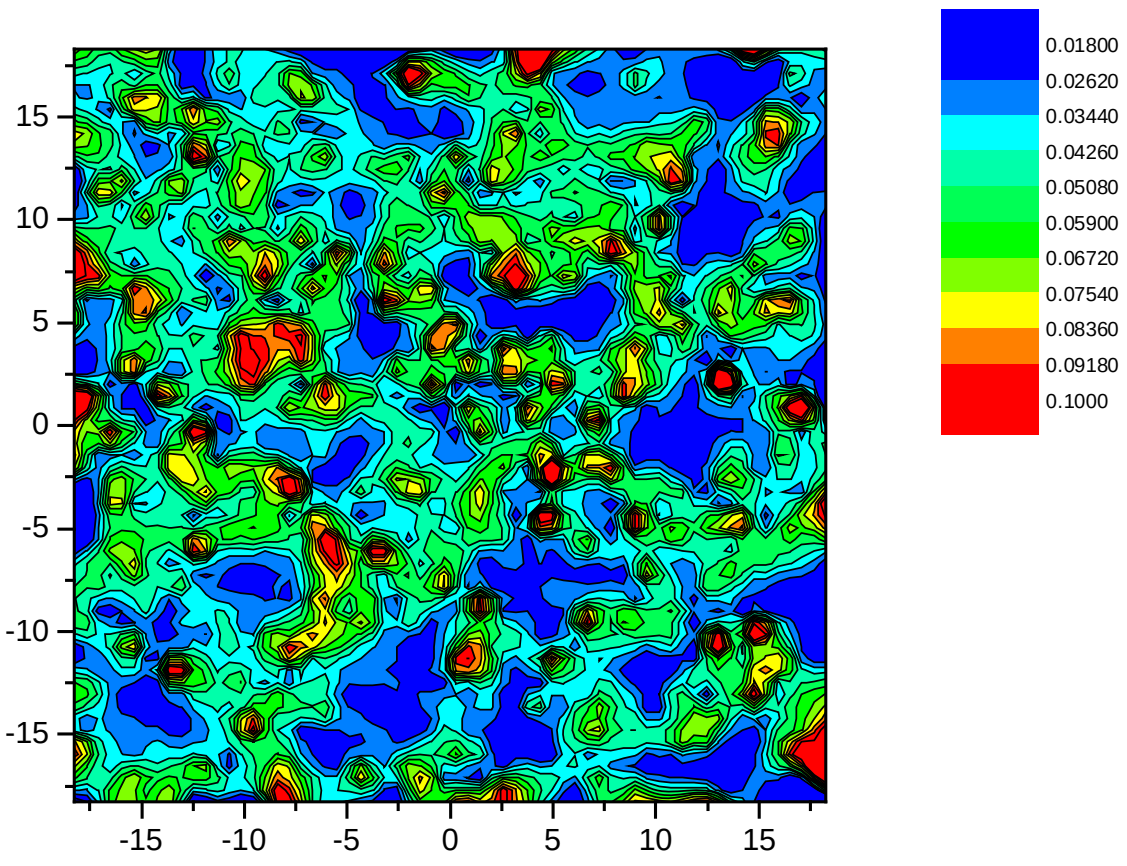

Figure 7. The spatial distribution of the relative free volume of the inherent structure for the same configuration used in Figs. 5 and 6. Many of the small isolated regions of high relative free volume (red) correspond to 'rattlers', particles trapped within a loose cage but unable to contribute to collective relaxation.

No significant correlation is evident. In spite of this general failure of the relative free volume to exhibit any strong correlation with the propensity, there is some cause for hope. Referring to the contour plot of the relative free volume in Fig. 7 (for the same configuration used in Figs. 5 and 6), we note the presence of a significant number of isolated particles with a high relative free volume. These 'rattlers' are the source of most of the 'false positives' in Fig.6. It seems reasonable to expect that the degree to which a particle can 'utilise' a neighbor's free volume depends upon the size of its own free volume. Based on this argument, we might be able to distinguish 'useful' free volume from that which cannot contribute to relaxation based on an analysis of clustering of particles with significant free volume. We are currently looking at such nonlocal approaches.

**5. Conclusions**

In this paper we have constructed the dynamic propensity of a configuration as an explicit measure of the degree to which the subsequent dynamics is governed by the configuration, at least as far out as the structural relaxation time. This real space map offers a complementary expression of what is referred to in the literature as landscape-dominated dynamics. Having established the degree to which the initial configuration does determine the spatial distribution of the propensity, we have sought to identify a reduced local measure of the configuration whose spatial distribution can provide a useful prediction of the propensity. While we found that high and low propensity was clearly associated with the small and large particles, respectively, some property other than local composition must also be involved. While the relative free volume of a particle exhibits little significant correlation with the propensity, it is noted that this result is influenced by the presence of 'rattlers', isolated particles with high relative free volume that do not appear to contribute to cooperative dynamics over the structural relaxation time.

With increasing supercooling comes the increasing collective character of particle motions. The degree to which a particle is constrained depends not just on the arrangement of neighbours but the degree to which those neighbours themselves are constrained, which in turn requires consideration of the neighbours' neighbours, and so on. This suggests that there might be a nonlocal extension of the relative free volume that reflects this cooperativity and distinguishes 'rattlers' from those particles whose free volume is available for collective reorganisation. This is the direction of our current work.